\documentclass{cpcauth}

\usepackage{epsfig}
\usepackage{amssymb}
\usepackage{graphicx}

\begin{document}

\begin{frontmatter}

\title{Detection of a small shift in a broad distribution}

\author{Bernd A.\ Berg}

\address{~~\\ Department of Physics, Florida State University, 
Tallahassee, FL 32306-4350, USA}

\date{May 31, 2012; revised Aug 27, 2014} 

\begin{abstract}
Statistical methods for the extraction of a small shift in broad data 
distributions are examined by means of Monte Carlo simulations. This 
work was originally motivated by the CERN neutrino beam to Gran Sasso
(CNGS) experiment for which the OPERA detector collaboration reported 
a time shift in a broad distribution with an accuracy of $\pm 7.8\,$ns,
while the fluctuation of the average time turns with $\pm 23.8\,$ns out 
to be much larger. Although the physical result of a big shift has been 
withdrawn, statistical methods that make an identification in a broad 
distribution with such a small error possible remain of interest.
\end{abstract}


\end{frontmatter}

\section{Introduction} \label{sec_intro}

\begin{figure*}[t] \begin{tabular}{c c}
\includegraphics[width=8cm]{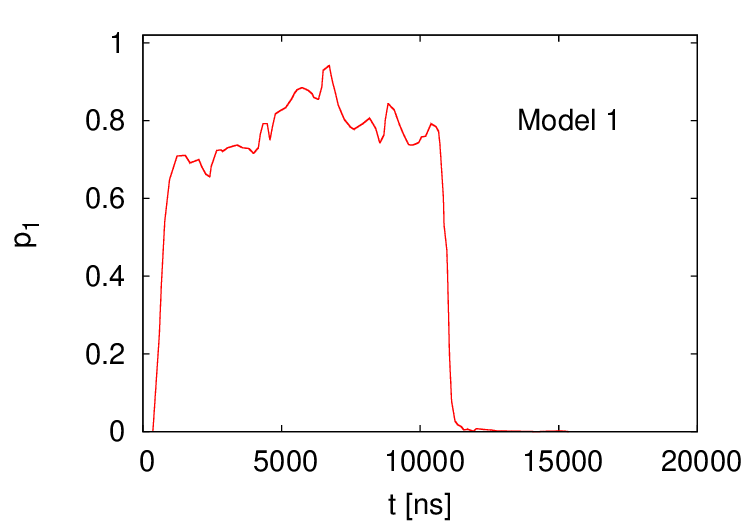} 
\hspace{0.5truecm}
\includegraphics[width=8cm]{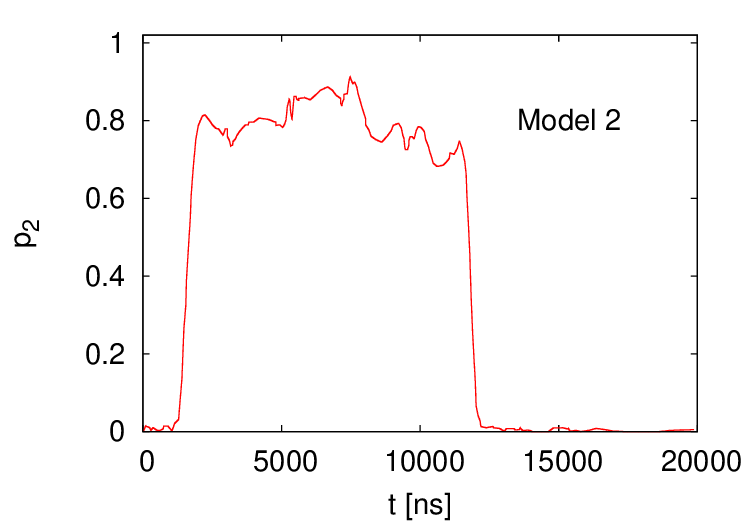} 
\end{tabular}
\caption{Departure time probability densities modeled after Fig.~11 
of Ref.~\cite{CNGS}.} \label{fig_2ppd}
\end{figure*}

In highly publicized CERN announcements \cite{CNGS} it was claimed that 
neutrinos from the CNGS arrived at Gran Sasso 
\begin{eqnarray} \label{deviation}
  \delta t=\rm \left[-57.8\pm 7.8\ (stat.)\,\matrix{+8.3\cr -5.9}\ 
                (sys.)\right]\, ns
\end{eqnarray}
too early, violating the $\delta t=0$ limit set by the speed of light. 
Meanwhile, initially overlooked systematic errors \cite{CNGS4} have 
wiped out the estimate of a large shift. But the estimate of the 
statistical error remains of interest as it exemplifies the extraction 
of a small shift from a broad distribution. The purpose of this 
article is to shed light on subtleties of an analysis, which leads 
to the statistical part of the estimate (\ref{deviation}). 

The CNGS sample of $15\,223$ neutrinos was produced in extractions 
that last about 10,500~ns each. Two different types of extractions
were used leading to probability densities (PD)
\begin{equation} \label{ppd}
  p_k(t)\,,~~k=1,2
\end{equation}
for neutrinos departure times, which are reproduced here in 
Fig.~\ref{fig_2ppd}. The PD used in our paper have been discretized 
in intervals of 1~ns and can be downloaded from the author's 
website~\cite{BBweb}. 

One can now perform a statistical bootstrap \cite{Efron} analysis 
by Monte Carlo (MC) generation of departure times with the PD of 
Fig.~\ref{fig_2ppd}. This is already remarked in \cite{CNGS}, 
where the application remains limited to testing of their maximum 
likelihood procedure on a sample of 100 MC data sets. As the MC 
generation of departure times can be repeated almost arbitrarily 
often with distinct random numbers, one can analyze and verify 
statistical methods that one wants to apply to discover a shift 
in the data.

For the uniform PD over 10$\,$500$\,$ns, it has been noted \cite{BH11} 
that with $n=16\,111$ events the 
variance of the departure time average 
$\overline{t}$ is approximately $\triangle\overline{t}=24\,$ns, i.e., 
much larger than the statistical error bar in Eq.~(\ref{deviation}). 
It will be discussed in this paper that the time shift $\delta t$ 
(\ref{deviation}) defined in \cite{CNGS} behaves indeed differently 
than a statistical fluctuation of the time average 
\begin{eqnarray} \label{otshift}
  \delta\overline{t} = \frac{n_1\,\delta\overline{t}_1+
  n_2\,\delta\overline{t}_2}{n_1+n_2}\,,~~
  \delta\overline{t}_i = \overline{t}_i - \widehat{t}_i\,,~~i=1,\,2\,.
\end{eqnarray}
Here $\overline{t}_i$ are the measured departure time averages,
$\widehat{t}_i$ are the mean departure times obtained from the 
underlying PD, and $n_i$ are the numbers of events in each 
extraction. The distinction between $\delta t$ (\ref{deviation}) 
and $\delta\overline{t}$ (\ref{otshift}) is made by an overline 
on $t$ or not. Obviously,
\begin{eqnarray} \label{means}
  \langle\delta t\rangle\ =\ \langle\delta\overline{t}\rangle 
\end{eqnarray}
holds for the expectation values, but their error bars behave differently.

To set the groundwork, it is shown in section~\ref{sec_uni} for the 
uniform distribution that a shift $\delta t = -57.8\,$ns can be 
identified with certainty (probability to miss it $<10^{-36}$) when 
there are 15$\,$223 events and the departure time range is $10\,500\,
$ns. In section~\ref{sec_MC} the MC generation of departure 
times is described. Section~\ref{sec_histo} gives examples of 
descriptive histograms from MC data. Suggested by the uniform 
distribution, the front tails of the distributions are of particular 
interest.  For their study the cumulative distribution function (CDF) 
is better suited than a histogram, because it allows easily to focus 
on outliers. This is investigated in section~\ref{sec_CDF}. To estimate 
the shift value $\delta t$, the maximum likelihood method is used in 
\cite{CNGS}. In section~\ref{sec_maxlikely} features of this method 
are calculated by applying it to a large number of MC generated 
departure time samples. 

Independently of the special example, the approaches discussed in 
sections~\ref{sec_uni} to~\ref{sec_maxlikely} are of interest, because 
they address the general problem of extracting a precise estimate of 
a shift from a broad distribution. Summary and conclusions follow in 
section~\ref{sec_sum}.

\section{Uniform Distribution \label{sec_uni} }

\begin{figure}[tb] \begin{center} 
\epsfig{figure=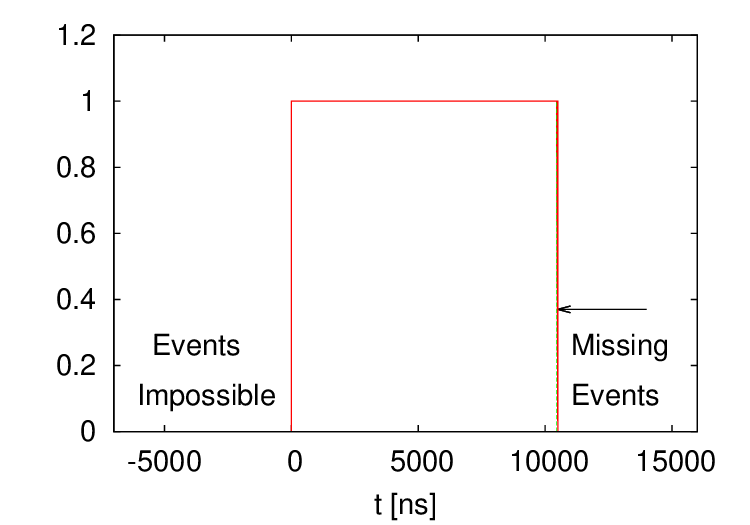,width=\columnwidth} 
\caption{Uniform distribution: Impossible (left) and missing events 
(in the enlarged thickness of the right border).  \label{fig_uni}} 
\end{center} \end{figure}

Using the uniform PD over a time window of 10$\,$500$\,$ns, the standard 
deviation of the average 
\begin{eqnarray} \label{average}
  \overline{t}\ =\ \frac{1}{n}\sum_{j=1}^n t^j
\end{eqnarray}
is for $n=15\,223$ events much larger than the statistical error bar 
quoted in Eq.~(\ref{deviation}), namely approximately
\begin{eqnarray} \label{statE}
  \triangle\overline{t}\ =\ 25\,{\rm ns}\,.
\end{eqnarray}
How can this be? That the average (\ref{average}) fluctuates with the 
variance (\ref{statE}) is unavoidable. However, the effect we are 
after is a systematic shift of each departure time by an amount 
$\delta t = -57.8\,$ns. Again for the uniform uniform distribution, 
drawn in Fig.~\ref{fig_uni}, it is easily illustrated that this can 
very well be identified. Events indicated on the left of the figure are 
impossible unless there is a shift. Now, with a shift of $-57.8\,$ns 
the probability to find a particular event to the left of the uniform 
PD is given by
\begin{eqnarray} \label{number}
  p\ =\ 57.8/10\,500\ =\ 0.005505\dots
\end{eqnarray}
and the probability to find none is
\begin{eqnarray} \label{likelihood}
  (1-p)^{15\,672}\ =\ 10^{-36.5}\,.
\end{eqnarray}
The distance of the smallest time from the left edge of the uniform
PD is a lower bound on $\delta t$ and a direct estimate for the 
time shift (\ref{deviation}) is
\begin{eqnarray} \label{deltuni}
  \delta t = n_{\rm left}\,10\,500\,{\rm ns}\,/\,15\,223\,,
\end{eqnarray}
where $n_{\rm left}$ is the number of events observed on the left
outside of the uniform PD. Confidence limits can be established
from the binomial distribution.

For the tiny range of $57.8\,$ns indicated by the somewhat 
thicker line on the right side of the uniform PD in Fig.~\ref{fig_uni}, 
the situation is the other way round. It has to be empty when there is
a shift by $\delta t = -57.8\,$ns. The probability that this happens by 
chance when there is in fact no shift is also given by (\ref{deltuni}).
When $\delta t$ is not known the distance of the largest measured time
from the right edge of the uniform PD is an upper bound on $\delta t$.

We do not pursue the uniform PD any further, because we are interested
in the more complicated case of the less sharp PD of Fig.~\ref{fig_2ppd}.

\begin{figure}[tb] \begin{center} 
\epsfig{figure=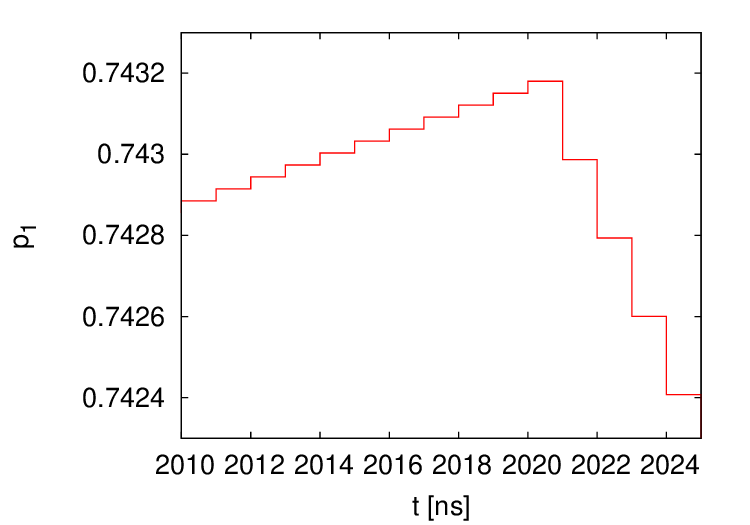,width=\columnwidth} 
\caption{Enlargement of the approximated PD for model~1 
over a small time region. \label{fig_d1mc}} 
\end{center} \end{figure} 

\section{MC generation of departure times \label{sec_MC}}

\begin{figure*}[t]   
\begin{tabular}{c c} 
\includegraphics[width=8cm]{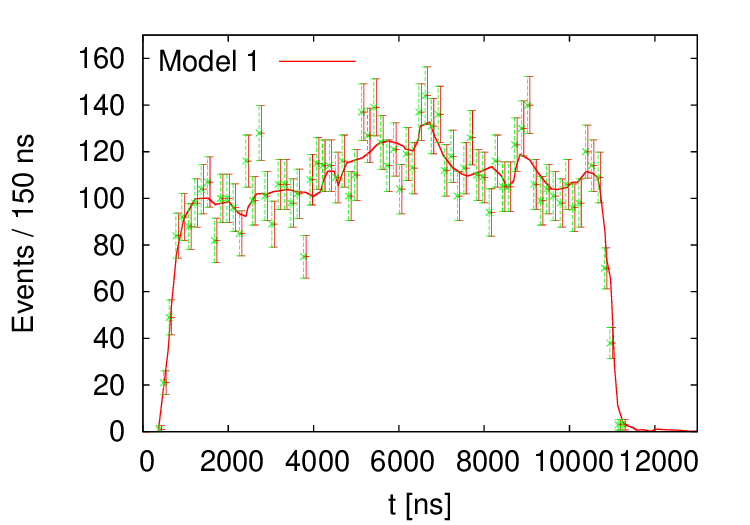}  
\hspace{0.5truecm}
\includegraphics[width=8cm]{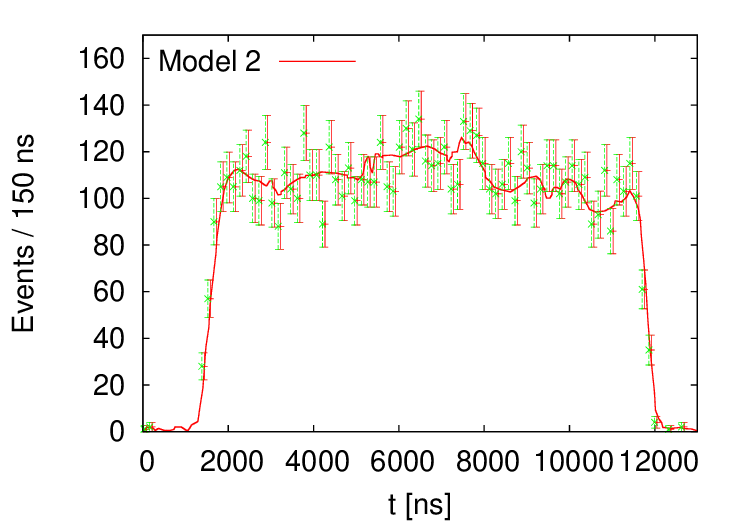}  
\end{tabular}
\caption{Typical histograms of departure times together with the PD of 
Fig.~\ref{fig_2ppd}. For the (online) green entries the generated time 
distribution is shifted by 57.8~ns to the left.} \label{fig_2h}
\end{figure*}

As mentioned in the introduction, the PD of Fig.~\ref{fig_2ppd} 
have been discretized in 1~ns intervals and are available on the 
Web~\cite{BBweb}. The resolution of 1~ns allows for easy MC generation 
of departure times and is sufficient for the intended accuracy 
of the estimate of a shift. In the following our thus defined models 
are labeled by $k=1,2$. The probabilities as function of time $t$ are 
defined by
\begin{eqnarray} \label{pkt}
  p_k(t)\ =\ p_k(i^t)~~{\rm for}~~ i^t\le t< i^t+1\,,
\end{eqnarray}
where $i^t$ are integers in ns units. As it is convenient for the MC 
generation of departure times, the normalization for the discretized 
PD is (distinct from Fig.~\ref{fig_2ppd}) chosen so that 
\begin{eqnarray} \label{Pmax}
  p^{\max}_k\ =\ \max_{i}\left[p_k(i)\right]\ =\ 1
\end{eqnarray}
holds. Proper normalizations $\sum_{i}p_k(i)\,\triangle t_k=1$
could still be achieved by choosing instead of ns some unconventional 
unit for $\triangle t_k$. For the generation of correctly distributed
random times this is irrelevant. For a short time range model~1 
probabilities $p_1(t)$ are enlarged in Fig.~\ref{fig_d1mc}.

After discretization the smallest $i^{\min}_k$ and largest $i^{\max}_k$ 
times with non-zero $p_k(i^t)$ values are 
\begin{eqnarray} \nonumber 
  i^{\min}_1 &=&  359~~~{\rm and}~~~i^{\max}_1\ =\ 15\,368\,,
  \\ \nonumber 
  i^{\min}_2 &=&~\,12~~~{\rm and}~~~i^{\max}_2\ =\ 19\,877\,.
\end{eqnarray} 
In particular for the large $i^t$ values, these ranges include a number 
of zero probabilities. MC generated departure times $t_k^j$ ($j=1,
\dots,n$) with $n$ the number of data are obtained from uniformly 
distributed random numbers $t$ in a range enclosing $\left(i^{\min}_k,
i^{\max}_k+1\right)$, here chosen to be (1,22000): for model~$k$ a 
proposed random time $t$ is accepted with 
probability $p_k(i^t)$ for $i^t\le t<i^t+1$. If $t$ is rejected, the 
procedure is repeated until a value gets accepted, which then becomes 
a data point $t_k^j$. Acceptance rates were close to 40\% and the 
random number generator of Ref.~\cite{Mar} has been used. To avoid 
direct hits of $i^{\min}_k$ and associated rounding problems, the 
proposed random times are shifted by $+2^{-25}$, which is half the 
discretization \cite{Bbook} of these random numbers.

\begin{figure*}[t]   
\begin{tabular}{c c} 
\includegraphics[width=8cm]{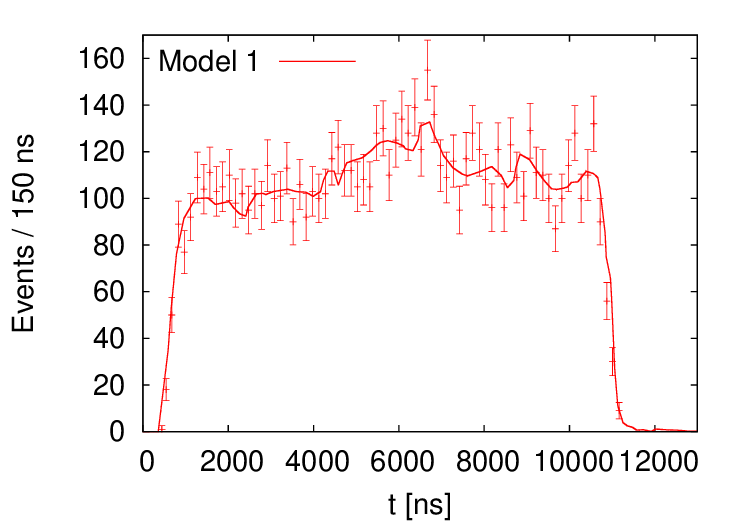}  
\hspace{0.5truecm}
\includegraphics[width=8cm]{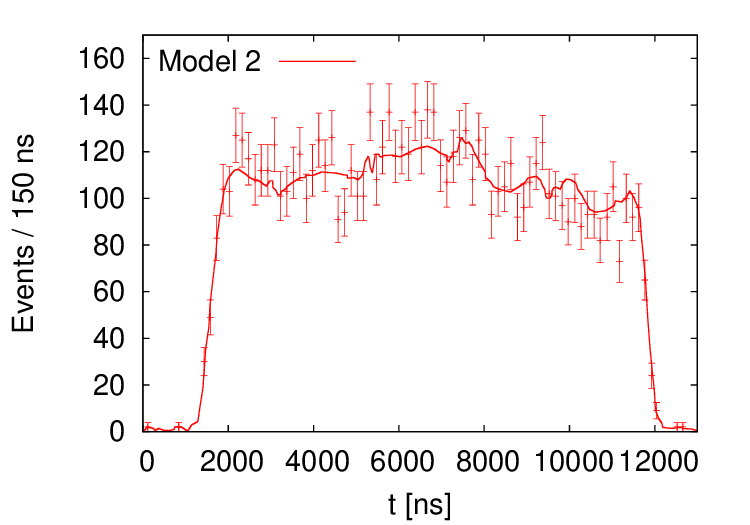}  
\end{tabular}
\caption{Histograms of departure times with a combined average 
fluctuation of $\delta\overline{t}=-59.6\,$ns.} \label{fig_2hf}
\end{figure*}

\begin{figure*}[t]   
\begin{tabular}{c c} 
\includegraphics[width=8cm]{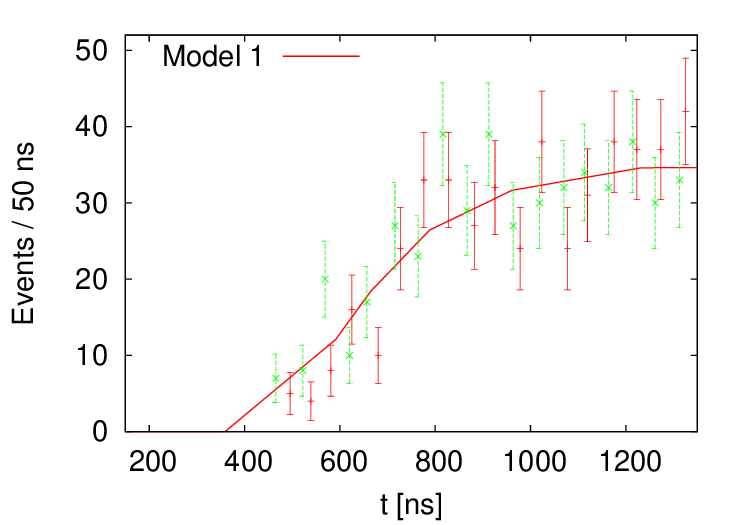}  
\hspace{0.5truecm}
\includegraphics[width=8cm]{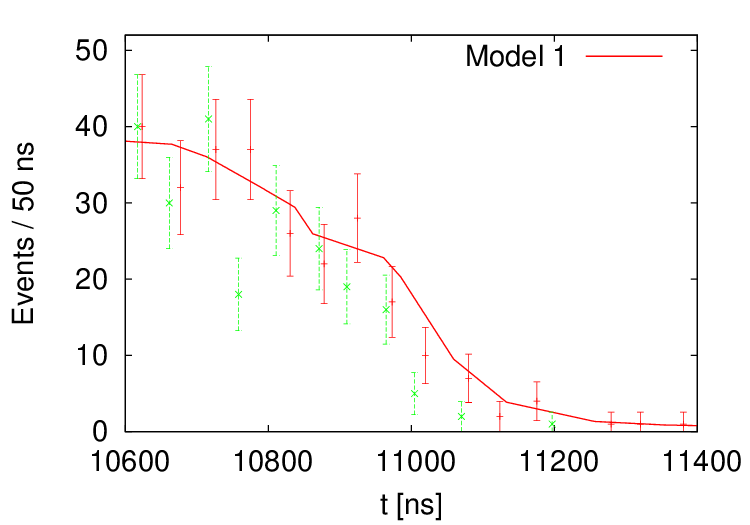}  
\end{tabular}
\caption{Model~1: Zoom of the leading (left plot) and trailing (right 
plot) MC generated departure time histograms from Fig.~\ref{fig_2hf}
(red online) with the PD and plotted together with Fig.~\ref{fig_2h} 
histograms shifted by $\delta t=-57.8\,$ns (green online).} 
\label{fig_h5s1} \end{figure*}

\begin{figure*}[t]   
\begin{tabular}{c c}
\includegraphics[width=8cm]{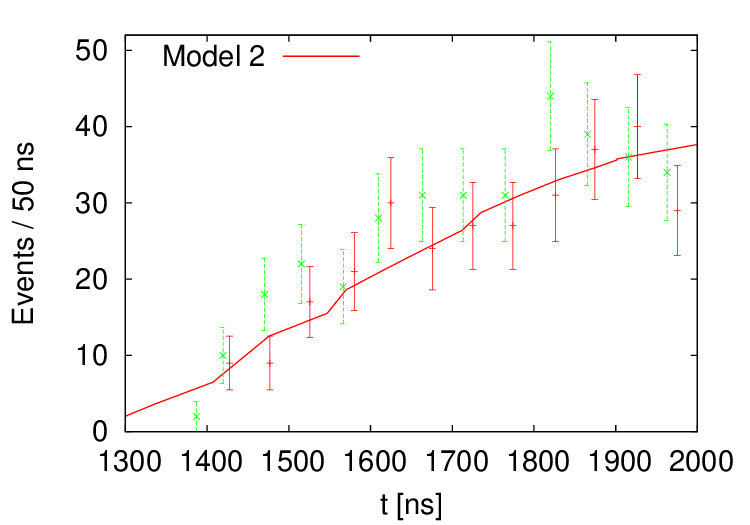}  
\hspace{0.5truecm}
\includegraphics[width=8cm]{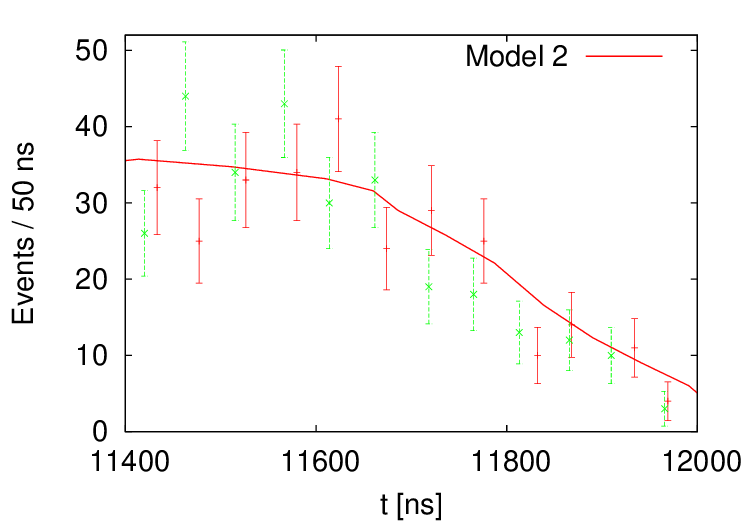}  
\end{tabular}
\caption{Model~2: Zoom of the leading (left plot) and trailing (right 
plot) MC generated departure time histograms from Fig.~\ref{fig_2hf}
(red online) with the PD and plotted together with Fig.~\ref{fig_2h} 
histograms shifted by $\delta t=-57.8\,$ns (green online).} 
\label{fig_h5s2} \end{figure*}

\section{Histograms \label{sec_histo} }

In this section we pursue descriptive statistics and show histograms 
from samples of $n=7\,612$ MC generated departure times per model. 
The exact departure time mean values for our discretized PD of 
Fig.~\ref{fig_2ppd} are (the hat indicates exact):
\begin{eqnarray} \label{tmean} 
  \hat{t}_1 = \langle t\rangle_1 = 5942.2\,{\rm ns}\,,\
  \hat{t}_2 = \langle t\rangle_2 = 6658.7\,{\rm ns}\,,
\end{eqnarray} 
where model~1 and~2 are again labeled by the corresponding subscripts.
We also know from these PD the exact standard deviations for time 
averages $\overline{t}_1$ and $\overline{t}_2$, each over $n=7\,612$
events:
\begin{eqnarray} \label{sigmatm1} 
  \widehat{\triangle}\overline{t}_1\ =\ 33.4\,{\rm ns}\,,~~~
  \widehat{\triangle}\overline{t}_2\ =\ 33.8\,{\rm ns}\,.
\end{eqnarray}
Their combined standard deviation is
\begin{eqnarray} \label{sigmatm} 
  \widehat{\triangle}\overline{t}\ =\ 23.8\,{\rm ns}\,,
\end{eqnarray}
which agrees almost with the estimate (\ref{statE}) from the uniform PD.

Together with the PD, Fig.~\ref{fig_2h} shows (red online) histograms 
of 150$\,$ns bin width for departure times from a typical~\cite{MC} 
MC generated sample of 7$\,$612 events per model and again the same 
histograms (green online) shifted by $-57.8\,$ns. This illustrates 
how small the effect is, which has to be unambiguously identified. 

Rounded to nearest integers the mean values for the time averages 
of the data of Fig.~\ref{fig_2h} are
\begin{eqnarray} \label{tm1} 
  \overline{t}_1 &=& (5936 \pm 34)\, {\rm ns~~(model~1)}\,,
  \\ \label{tm2}
  \overline{t}_2 &=& (6703 \pm 34)\, {\rm ns~~(model~2)}\,.
\end{eqnarray}
Both are within statistical uncertainties consistent with the mean 
values (\ref{tmean}) of the PD. The actual statistical fluctuations
of these two MC data sets are (using $\overline{t}_1=5935.5$ and 
$\overline{t}_2=6703.2$)
\begin{eqnarray} \label{delta1} 
  \delta\overline{t}_1 &=& \overline{t}_1-\hat{t}_1\ 
  =\ - 6.7\,{\rm ns}\,,\\ 
  \delta\overline{t}_2 &=& \overline{t}_2-\hat{t}_2\ 
  =\ 44.5\,{\rm ns}\,,
\end{eqnarray}
with an average of 18.9$\,$ns.

To compare the shift of $\delta t = -57.8\,$ns with a statistical 
fluctuation of at least the same size, the MC generation was repeated 
one million times and in this process 7$\,$608 samples (0.76\%) with 
statistical fluctuations $\delta\overline{t}\le -57.8\,$ns were 
encountered.  The PD of the first of these MC samples are shown in 
Fig.~\ref{fig_2hf}. They feature the time averages
\begin{eqnarray} \label{tm3} 
  \overline{t}_1 &=& (5907 \pm 34)\,{\rm ns,~~(model~1)}\,,
  \\ \label{tm4}
  \overline{t}_2 &=& (6575 \pm 34)\,{\rm ns,~~(model~2)}\,,
\end{eqnarray}
with the actual statistical fluctuations (using $\overline{t}_1=5906.7$ 
and $\overline{t}_2=6575.0$)
\begin{eqnarray} \label{delta3a} 
  \delta\overline{t}_1 &=& \overline{t}_1-\hat{t}_1\ 
  =\ -35.5\,{\rm ns}\,,\\  \label{delta3b}
  \delta\overline{t}_2 &=& \overline{t}_2-\hat{t}_2\ 
  =\ -83.7\,{\rm ns}\,,
\end{eqnarray}
resulting in an average of $-59.6\,$ns. \smallskip

Figures~\ref{fig_h5s1} and \ref{fig_h5s2} zoom into the leading and 
trailing parts of the departure times of Fig.~\ref{fig_2hf} (red online) 
and compare them with the shift by $-57.8$ of Fig.~\ref{fig_2h} (green 
online). The binsize is from $150\,$ns reduced to $50\,$ns, so that it 
is now slightly smaller than the shift, while the noise in the histogram 
increases. By visual inspection of histograms it is difficult to 
distinguish a shift of all times from a statistical fluctuation. The 
analysis of the uniform distribution in section~\ref{sec_uni} suggests 
to display the tails of the distribution event by event. Using the 
empirical CDF this is done next.

\begin{figure*}[t] 
\begin{tabular}{c c}
\includegraphics[width=8cm]{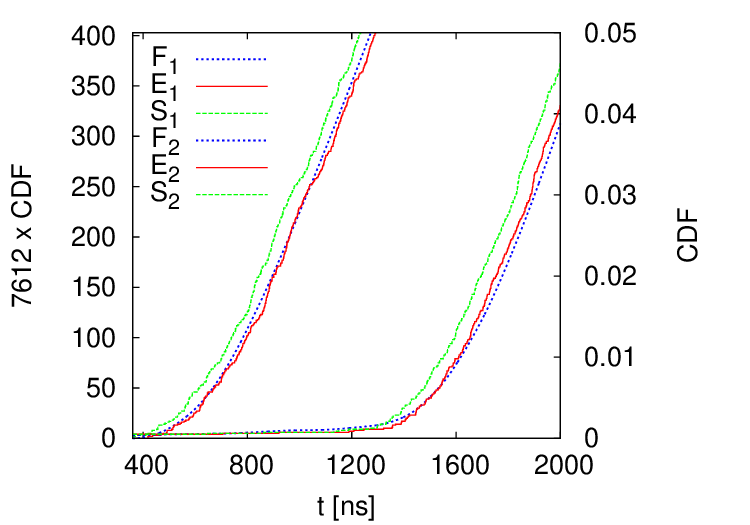}  
\hspace{0.5truecm}
\includegraphics[width=8cm]{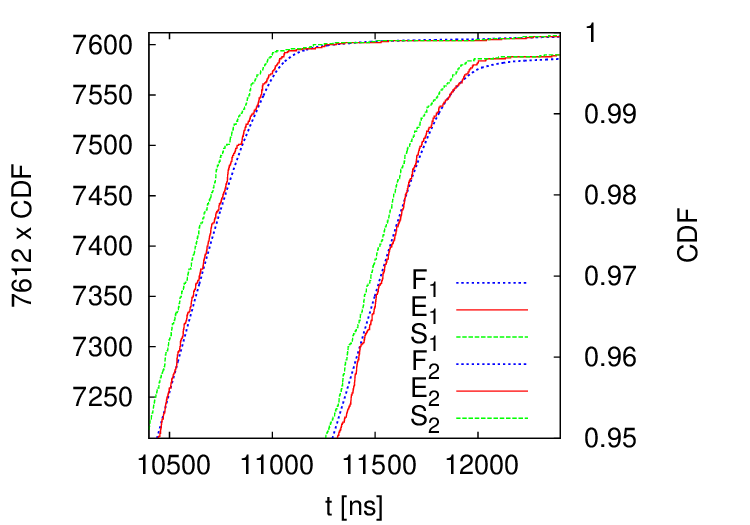}  
\end{tabular}
\caption{Tails of the CDF ($F_k$ exact, $E_k$ empirical, $S_k$ 
empirical shifted by $-57.8\,$ns, $k=1,2$ models).} \label{fig_c8d1}
\end{figure*}

\begin{figure*}[t] 
\begin{tabular}{c c}
\includegraphics[width=8cm]{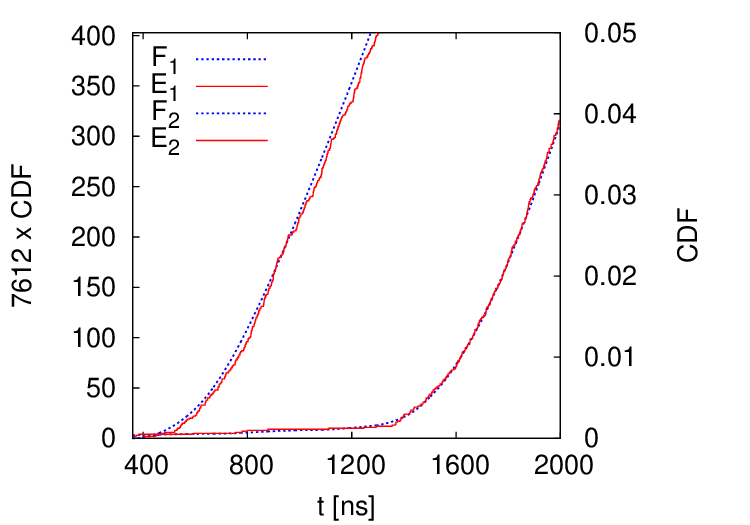}  
\hspace{0.5truecm}
\includegraphics[width=8cm]{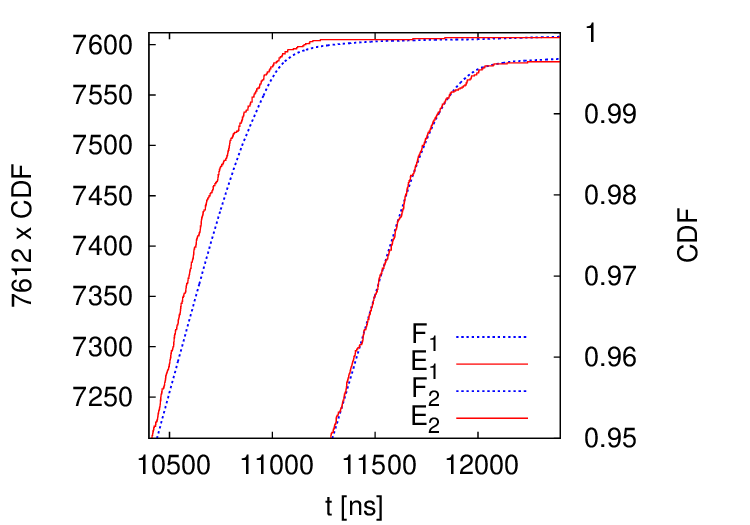}  
\end{tabular}
\caption{Tails of the CDF for departure times with a $-59.6\,$ns 
statistical fluctuation ($F_k$ exact, $E_k$ empirical).} 
\label{fig_c8d3}
\end{figure*}

\begin{figure}[tb] \begin{center} 
\epsfig{figure=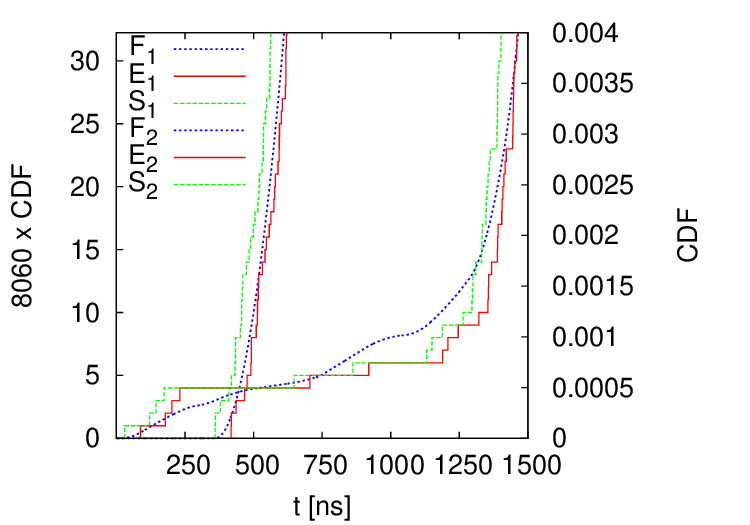,width=\columnwidth} 
\caption{Enlargement of the left part of Fig.~\ref{fig_c8d1}.
\label{fig_elc8d1l}} 
\end{center} \end{figure} 

\section{Cumulative distribution functions (CDF)} \label{sec_CDF}

For continuous distributions a problem of histograms is that a
binsize needs to be chosen. In our case we need a good resolution 
on the time axis, i.e., a small binsize. However, this increases 
the noise of the bin average. As the empirical CDF has no such free 
parameter, it is better suited for the analysis at hand. In particular,
it is well suited for a display of the tails of a distribution. Based 
on the binomial distribution we will develop their quantitative analysis 
relying on uniformly distributed random variables, similarly as for the 
goodness of fit~\cite{Bbook}.

The exact CDF of our PD models are
\begin{equation} \label{Fk}
  F_k(t)\ =\ \int_{-\infty}^t dt'\,p_k(t')\,,~~k=1,2\,,
\end{equation}
where $p_k(t)$ is given by (\ref{pkt}). The empirical CDF $E_k(t)$
are calculated from the data (see, e.g., \cite{Bbook}), which are 
here MC generated random times. Let them be $t^1_k,\dots,t^n_k$ and 
$t^{\pi_1}_k,\dots,t^{\pi_n}_k$ sorted, so 
that $t^{\pi_i}_k< t^{\pi_{i+1}}_k$ holds for $i=0,\dots,n+1$, where 
the definitions $t_k^{\pi_0}=-\infty$ and $t_k^{\pi_{n+1}}=+\infty$ 
have been added. The empirical CDF are then defined by
\begin{eqnarray} \label{Ek}
  E_k(t) &=& {i\over n}~~~{\rm for}~~~t^{\pi_i}_k\le t<t^{\pi_{i+1}}_k,
  \\ \nonumber i &=& 0, 1,\dots , n , n+1\,.
\end{eqnarray} 
\noindent Shifted by $\delta t$ the empirical CDF are
\begin{equation} \label{Sk}
  S_k(t)\ =\ E_k(t-\delta t)\,.
\end{equation}
For the data sets of random times used in the previous section 
$F_k(t)$, $E_k(t)$ and $S_k(t)$ with $\delta t=-57.8\,$ns are displayed 
in Fig.~\ref{fig_c8d1}. The scale on the right ordinates gives the usual 
probability definition for which a CDF is in the range $[0,1]$, while 
the scale on the left ordinates is chosen to count the number of 
included data points. Fig.~\ref{fig_c8d1} should be compared with 
Fig.~\ref{fig_c8d3}, which features the statistical fluctuation 
(\ref{delta3a}), (\ref{delta3b}) of the average time by $-59.6\,$ns. 
While the $-57.8\,$ns shift is clearly visible for all four cases of 
Fig.~\ref{fig_c8d1}, there is ony in one case a similar signal in 
Fig.~\ref{fig_c8d3}.

In the figures on the right side $F_k(t)=1$ is only slowly approached, 
because in the PD there are small contributions from the region starting 
between $15\,000\,$ns and $20\,000\,$ns (about 30 events for model~2 for 
which the effect is larger than for model~1). This does not spoil the 
over-all picture, so that the accuracy of the PD in this region does 
not really matter.

Visually, shift and statistical fluctuation behave already differently,
though not always. The task is now to develop quantitative criteria, 
which are in this section based on the binomial distribution. Defining
\begin{equation} \label{Fik}
  F^i_k\ =\ F_k(t^{\pi_i}_k)~~{\rm and}~~G^i_k=1-F^i_k
\end{equation}
the probability for a single data point $t$ to be in the range $0\le 
F_k(t) \le F^i_k$ is $F^i_k$ and its probability to be in the range 
$F^i_k< F_k(t)\le 1$ is $G^i_k$. The probability that $i$ out of $n$ 
data points are in the $0\le F_k(t) \le F^i_k$ range is 
\begin{equation} \label{pik}
  p^i_k\ =\ \frac{n!}{(n-i)!\,i!}\,\left(F^i_k\right)^i\,
            \left(G^i_k\right)^{n-i}\,.
\end{equation}
If there is a shift by $\delta t<0$ we expect that more data points 
populate the $0\le F_k(t) \le F^i_k$ range, so that
\begin{equation} \label{Qik}
  Q^i_k\ =\ \sum_{j=i}^n p^j_k
\end{equation}
becomes small. It is easy to see that the $Q^i_k$ are uniformly 
distributed random variables, when the hypothesis is correct that
the data points are created with the distribution $F_k(t)$. So the 
usual interpretation \cite{Bbook} of a likelihood that the discrepancy 
between the hypothesis and the data is due to chance applies to the $Q^i_k$.


To present an example we give in Fig.~\ref{fig_elc8d1l} an enlargement 
of the left tail from Fig.~\ref{fig_c8d1} by picking out the 32 smallest 
times per model. Upon inspection of the numbers one finds for model~1 
$t_1^{\pi_1}=359.985\,$ns and $F^1_1=F_1(t_1^{\pi_1})=5.7\times 
10^{-9}$. This implies $Q^1_1=4.3\times 10^{-5}$ for model~1. 
For model~2 $t^{\pi_1}_2=29.968\,$ns, which gives $F^1_2=F_2(t_2^{\pi_1})
=3.5\times 10^{-6}$ and for $n=7612$ one finds $Q^1_2=0.026$. Combined, 
these two events alone give a probability $Q=Q^1_1\,Q^1_2 = 1.15\times
10^{-6}$ for the likelihood that the discrepancy is due to chance.

\begin{figure}[tb] \begin{center} 
\epsfig{figure=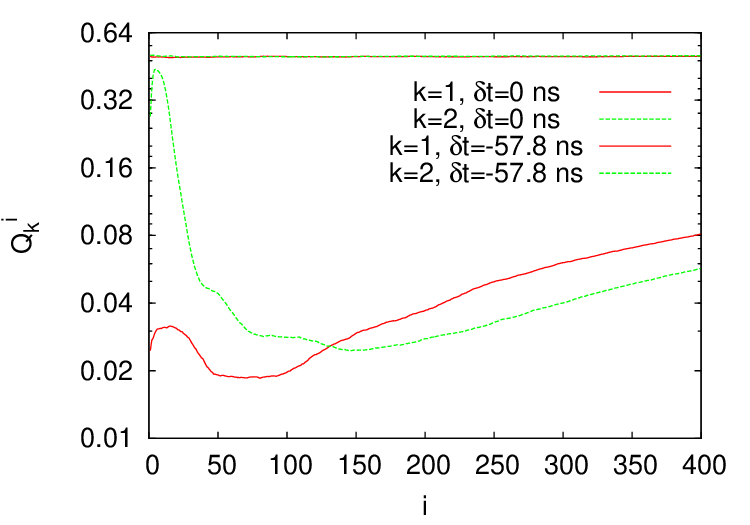,width=\columnwidth} 
\caption{Average $Q^i_k$ as functions of $i$ from $10\,000$ 
samples. \label{fig_Qik} } 
\end{center} \end{figure} 

To understand the typical behavior encountered for our two distribution 
functions, we generated $10\,000$ MC samples. In Fig.~\ref{fig_Qik} 
average values for $Q^i_k$ are shown using a log scale on the ordinate. 
The almost straight lines at 0.5 are the averages obtained when one 
generates random times with the underlying PD. The two other lines 
correspond to the average $Q^i_k$ obtained when each sample undergoes 
a shift of $-57.8\,$ns. For increasing $i$ (out of the range of the
figure) they will approach 0.5, because the shift has little effect
in the middle of a broad distribution.

At a first glance the average $Q^1_k$ values for the smallest times
look far less promising than what we may have hoped for from the
inspection of our trial sample. For model~1 the average is at 0.025
and for model~2 even at 0.27. At a second look we shall see that
this comes mainly from the outliers of the non-Gaussian distribution
of the $Q^1_k$, while for a majority of cases our trial sample
remains typical. Besides $Q^1_k$ for the smallest times, an 
attractive choice for $Q^i_k$ is at or close to the minimum of
the average, which we choose to be $i=80$ for model~1 and $i=160$
for model~2. Due to the increased number of data there are fewer
outliers than for $Q^1_k$. Note that for different $i$ values the 
$Q^i_k$ are not statistically independent, because the calculation 
of $Q^i_k$ includes all data already used for $Q^j_k$ with $j<i$.

\begin{figure}[tb] \begin{center} 
\epsfig{figure=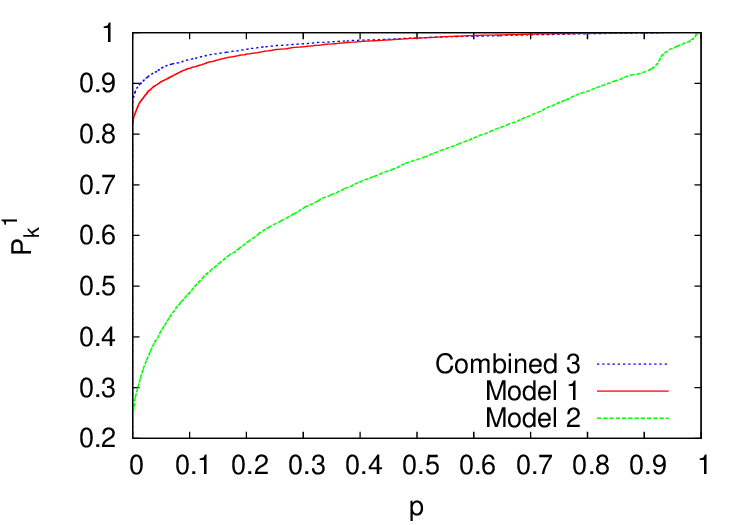,width=\columnwidth} 
\caption{Probabilities $P^1_k$ for $Q^1_k\le p$, $\delta t=-57.8\,$ns,
from 10$\,$000 samples.  \label{fig_p01} } 
\end{center} \end{figure} 

We define $P^i_k(p)$ to be the probability for $Q^i_k\le p$.
Fig.~\ref{fig_p01} plots $P^1_k(p)$ and we find $Q^1_1=0$
for more than 80\% of the samples and $Q^1_2=0$ for almost
23\% of the samples. To combine the results from both models 
we notice that
\begin{equation} \label{Qi3}
  Q^{ij}_3\ =\ Q^i_1\,Q^j_2\,\left[1-\ln(Q^i_1\,Q^j_2)\right]
\end{equation}
is again a uniformly distributed random variable with the
interpretation that the discrepancy between data and hypothesis is due 
to chance ($Q^{ij}_3\to Q^i_1Q^j_2$ for $Q^i_1$ or $Q^j_2$ to zero). We 
use the notation $Q^i_3=Q^{ii}_3$ and find $Q^1_3=0$ for more than 85\% 
of the samples, where (\ref{Qi3}) is of course performed before sorting.

\begin{figure}[tb] \begin{center} 
\epsfig{figure=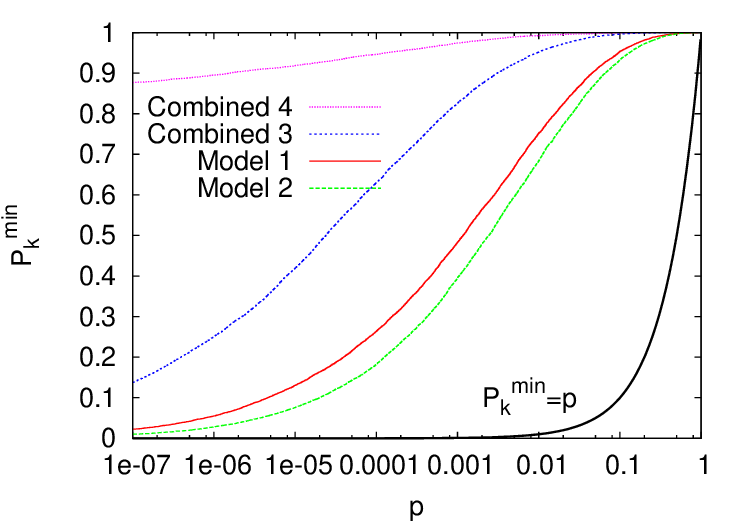,width=\columnwidth} 
\caption{Probabilities $P^{\min}_k$ for $Q^{\min}_k\le p$, $\delta 
t=-57.8\,$ns, from 10$\,$000 samples (log scale on the abscissa). 
\label{fig_1pmin} } 
\end{center} \end{figure} 

We define now $Q^{\min}_1=Q^{80}_1$, $Q^{\min}_2=Q^{160}_2$ and 
$Q^{\min}_3$ the combination (\ref{Qi3}) of the two. $P^{\min}_k(p)$ 
are the corresponding probabilities for $Q^{\min}_k\le p$. In 
Fig.~\ref{fig_1pmin} the $P^{\min}_k(p)$ are plotted versus $p$ 
using a log scale on the abscissa (the line $P^{\min}_k=p$ is no
longer straight). The $Q^{\min}_k$ probabilities 
tend to be small, but never zero. For the combined data we find 
$Q^{\min}_3\le 10^{-7}$ for about 14\% and $Q^{\min}_3\le 10^{-3}$ 
for more than 82\% of the samples. Now $Q^{\min}_3$ works quite 
independently from $Q^1_k$, reducing the situations, where their 
combined estimate $Q^{\min}_4=Q^{\min}_3\,Q^1_3\,\left[1-\ln(Q^{\min}_3
\,Q^1_3)\right]$ gives $Q^{\min}_4>10^{-3}$ to approximately 2.6\%. 
Compare Fig.~\ref{fig_1pmin}. Although the application of 
Eq.~(\ref{Qi3}) is not entirely correct for $Q^{\min}_4$, the related 
bias is small as it comes from the contribution of single data points 
$t^{\pi_1}_k$ to each of the $Q^{\min}_k$.

\begin{figure}[tb] \begin{center} 
\epsfig{figure=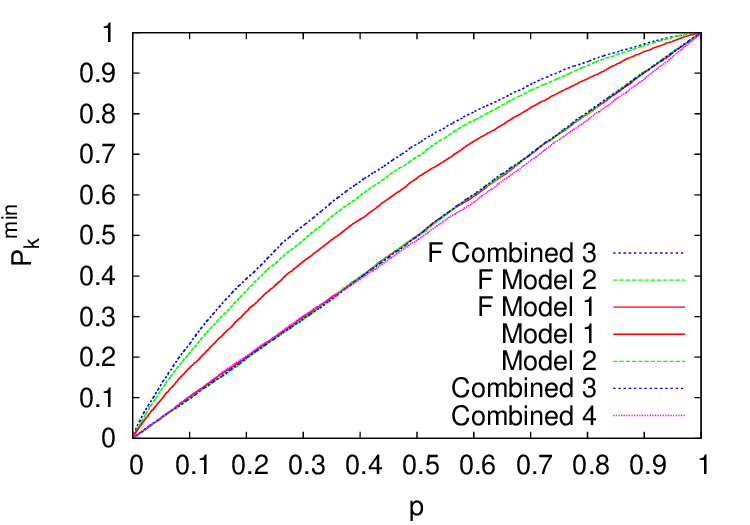,width=\columnwidth} 
\caption{Diagonal lines: Probabilities $P^{\min}_k$ for 
$Q^{\min}_k\le p$, $\delta t=0\,$ns. Upper curves (F): The same with
a statistical fluctuation $\delta\overline{t}\le -57.8\,$ns of
each sample average. Each curve relies on 10$\,$000 samples. \label{fig_2pmin} } 
\end{center} \end{figure} 

\begin{figure}[tb] \begin{center} 
\epsfig{figure=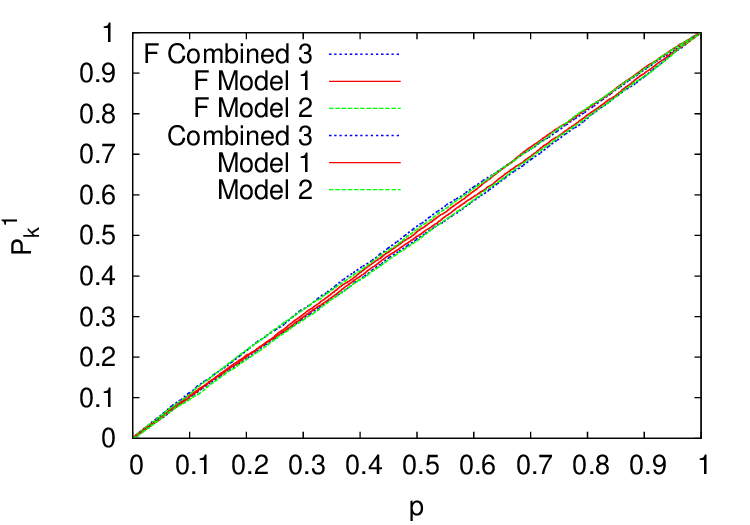,width=\columnwidth} 
\caption{Probabilities $P^1_k$ for $Q^1_k\le p$, $\delta t=0\,$ns
and (F) the same with a statistical fluctuation $\delta\overline{t}
\le -57.8\,$ns of each sample average. Each curve relies on 
10$\,$000 samples. \label{fig_2p01} } 
\end{center} \end{figure} 

For the peace of mind we show in Fig.~\ref{fig_2pmin} the $P^{\min}_k$
values when no shift is applied to the data. As claimed the $Q^{\min}_k$
turn out to be uniformly distributed random variables in the $[0,1)$ 
range with a small bias visible for $Q^{\min}_4$. These are the almost 
straight $P^{\min}_k(p)=p$ lines. The upper part of the figure shows 
$P^{\min}_k(p)$ from $10\,000$ samples, which are selected so that each 
exhibits a statistical fluctuation $\delta\overline{t}\le -57.8\,$ns
(as about 0.76\% of the samples exhibit such a fluctuation it requires 
the generation of more than 13 million samples). In contrast to 
Fig.~\ref{fig_1pmin} the $P^{\min}_k(p)$ probabilities for $Q^{\min}_k
\le p$ are only slightly enhanced when compared to the straight line. 
The same analysis shows for $Q^1_k$ practically no deviation from the 
straight $P^1_k(p)=p$ lines as is demonstrated in Fig.~\ref{fig_2p01}. 
These statements remain true when a bias which enhances the initial 
tail of the PD (discussed in \cite{BH11}) contributes to the shift 
in the mean value $\delta\overline{t}$.

In summary, there is high probability that a study of the initial 
tails of the departure time distributions reveals whether there 
is a statistically relevant shift or not. A similar study could 
be performed for the back tail of the departure times. One could 
further expand the analysis of this section to provide an explicit 
estimate of the shift $\delta t$, similarly as with Eq.~(\ref{deltuni}) 
for the uniform PD, but now by bootstap simulation. Here we leave such 
an estimate to using the maximum likelihood method of the next section.

\begin{figure*}[t] 
\begin{tabular}{c c}
\includegraphics[width=8cm]{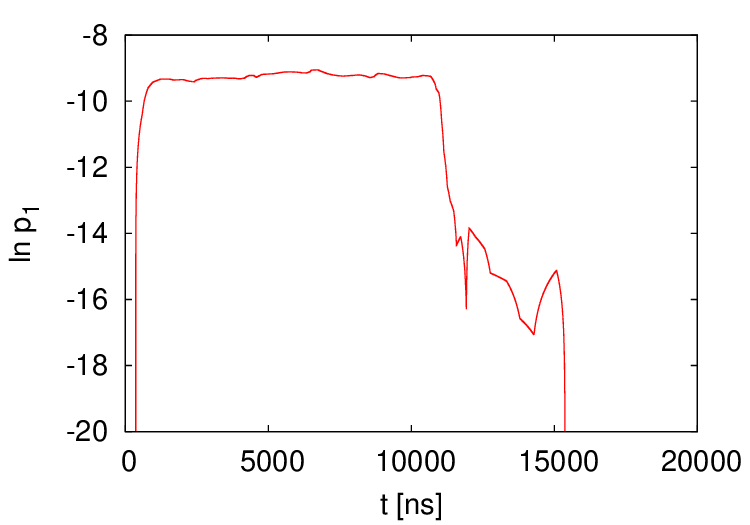} 
\hspace{0.5truecm}
\includegraphics[width=8cm]{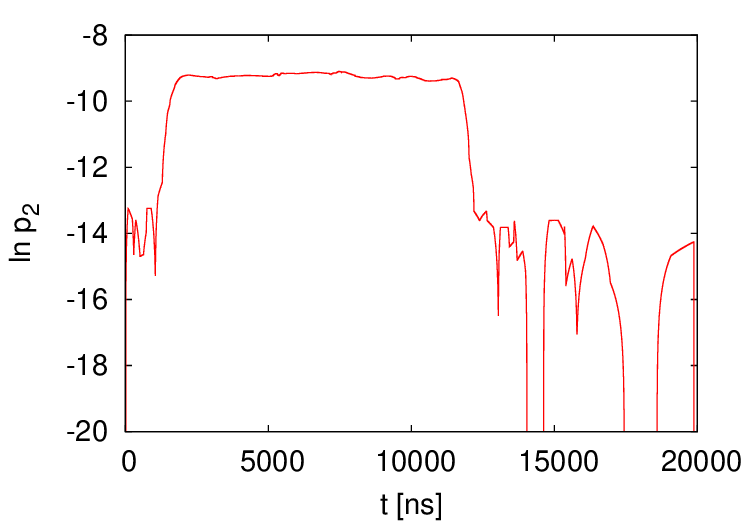} 
\end{tabular}
\caption{Departure time probability densities of Fig.~\ref{fig_2ppd} 
on a log scale.} \label{fig_ln2ppd}
\end{figure*}

\begin{figure*}[t] 
\begin{tabular}{c c}
\includegraphics[width=8cm]{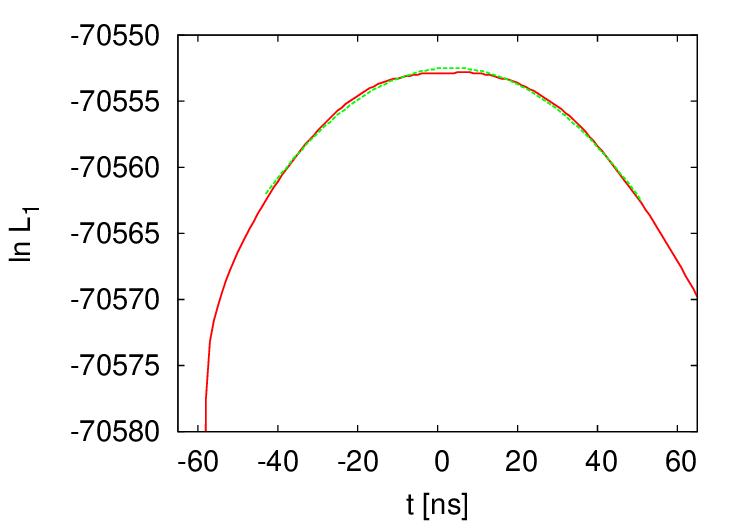} 
\hspace{0.5truecm}
\includegraphics[width=8cm]{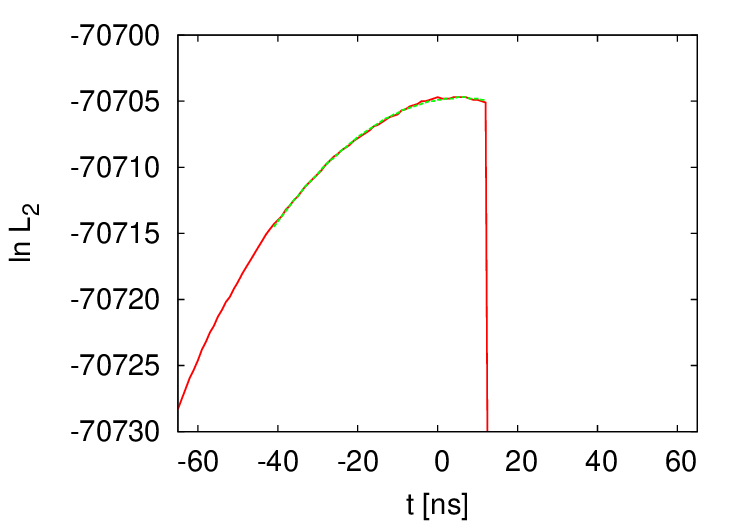} 
\end{tabular}
\caption{Log-likelihood functions (red online) of our typical sample 
for model~1 and~2. Parabolic fits (green online) are indicated for 
the central regions.} \label{fig_lnw}
\end{figure*}

\begin{figure}[tb] \begin{center} 
\epsfig{figure=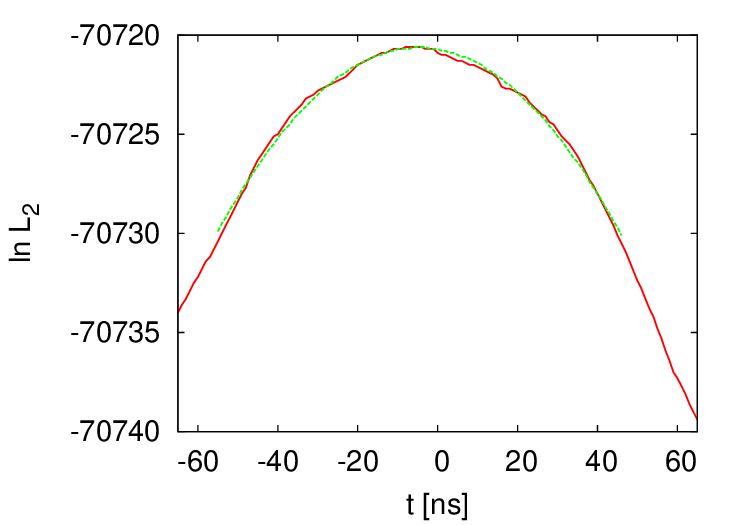,width=\columnwidth} 
\caption{Log-likelihood function (red online) for model~2 from one
of our first ten samples. A parabolic fit (green online) is indicated 
for the central region.}  \label{fig_7lnw} 
\end{center} \end{figure} 

\begin{figure}[tb] \begin{center} 
\epsfig{figure=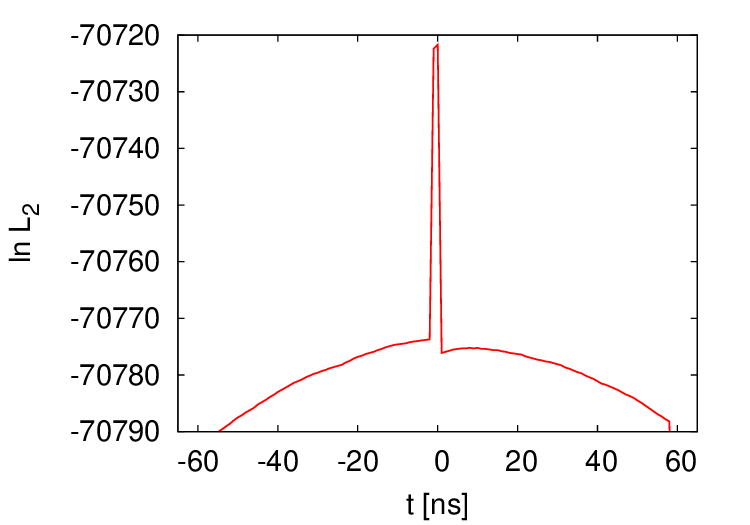,width=\columnwidth} 
\caption{Log-likelihood function (red online) for model~2 from a
sample for which a parabolic fit does not work. \label{fig_161201lnw} }
\end{center} \end{figure} 

\section{Maximum likelihood method} \label{sec_maxlikely}

For the statistical estimate of the time shift value $\delta t$ a 
maximum likelihood method is the choice of Ref.~\cite{CNGS}. Its 
properties are studied in this section using $n=10\,$000 MC generated 
samples for each of the PD. 

The log likelihoods for our PD (\ref{ppd}) are given by
\begin{equation} \label{lnLk}
  \ln L_k(t)\ =\ \sum_{i=1}^n\ln\left[p_k(t^i+t)\right],\ k=1,2
\end{equation} 
where the times $t^i$ are MC generated with the PD $p_k(t)$. It is 
instructive to replot Fig.~\ref{fig_2ppd} for logarithms of the 
probabilities. This is done in Fig.~\ref{fig_ln2ppd} with $p_k(t)$
normalized to $\sum_i p_k(t^i)=1$, where (in accordance with our 
discretization) $t^i$ is incremented in 1~ns steps. These plots exhibit 
clearly the locations of small and zero probabilities, which data will 
(statistically) avoid when they are created with the underlying PD 
$p_k(t)$. However, when shifting the $t^i$ values in Eq.~(\ref{lnLk}) 
by $t$, regions of low probabilities may be hit as discussed in the 
previous sections. 

Our samples are generated without a shift $\delta_{\rm shift} t$, 
because the only effect of adding it would be that all $\delta t$
reported in this section become transformed according to
\begin{equation} \label{shift}
  \delta t\ \to\ \delta t + \delta_{\rm shift} t\,.
\end{equation} 
Consequently, the $\delta t$ discussed in this section deal with the 
various error sources.

For our typical sample (see section~\ref{sec_histo}) \cite{MC}, the 
resulting $\ln L_k(t)$ functions 
are displayed in Fig.~\ref{fig_lnw}. For model~1 we see for decreasing 
$t$, just before $-60\,$ns, a sharp drop $\ln L_1(t)\to-\infty$. This 
comes because the smallest time, $t^{\pi_1}+t$ hits a tiny probability, 
which we already encountered in the previous section. For model~2 the 
log-likelihood function $\ln L_2(t)$ is smooth down to $t=-65\,$ns, 
while for positive $t>12\,$ns the largest data point $t^{\pi_n}+t$ 
hits a region of very small probability.

Fig.~\ref{fig_7lnw} shows for model~2 another log-likelihood function,
$\ln L_2(t)$, from our first ten samples and this plot is quite similar 
to the one for extraction~2 in \cite{CNGS}. In the samples one finds a 
variety of deviations from Gaussian shapes, small ones due to the bulk 
structure of the PD and large ones due to a few events in the tails. 

Following \cite{CNGS} a step further, parabolic fits are made to the 
central region of each sample. To automatize the fitting procedure 
the central region of a sample was defined to be the 
$t$ range connected with $\delta t_k$ so that $\ln L_k(t) > \ln 
L^{\max}_k-10$ holds, where the maximum of the log-likelihood 
function is
\begin{equation} \label{Lmax}
  \ln L^{\max}_k= \max_t\left[\ln L_k(t)\right]= \ln L_k(\delta t_k)\ .
\end{equation}
Such parabolic fits are included in Fig.~\ref{fig_lnw} and~\ref{fig_7lnw}.

 Let us investigate $\delta t_k$ estimates for random times generated with 
the PD (\ref{pkt}). In the limit of infinite statistics
\begin{equation} \label{Lmax0}
  \lim_{n\to\infty} \frac{1}{n}\,\ln L_k(t) = 
  \sum_i p_k(i)\,\ln p_k(i+t)
\end{equation}
holds, where the sums corresponds to our discretization time steps of 
1~ns. They include $p_k(i)=0$ contributions. For $t=0$ they are zero 
due to 
\begin{equation}
  \lim_{p\to 0} \left(p\,\ln p\right)\ =\ 0\,.
\end{equation}
For $t\ne 0$, e.g.\ already $t=\pm 1\,$ns, we can create mismatched 
contributions $p_k(i)\ln p_k(t+t)$ with $p=p(i)$ finite and 
$q=p(i+t)=0$, so that 
\begin{equation}
  \lim_{q\to 0} \left(p\,\ln q\right)\ =\ -\infty
\end{equation}
becomes possible. Hence, in the limit of infinite statistics
\begin{equation}
  \lim_{n\to\infty}\left\langle\delta t_k\right\rangle_n\ =\ 0\
   =\ \lim_{n\to\infty}\left\langle\left(\delta t_k\right)^2
  \right\rangle_n\,, 
\end{equation}
where $\langle\dots\rangle_n$ are expectation values for samples of 
$n$ data per model. Deviations of $\delta t_k$ from zero reflect 
statistical fluctuations and bias due to a finite statistics. In the 
following we investigate this for $n=7\,612$.

Direct estimates of $\delta t_k$ are obtained by simply evaluating the 
log-likelihood functions (\ref{lnLk}) for a sufficiently large region
around its maximum, which is here taken to be [$-65\,$ns,$65\,$ns]. The 
positions of the maxima of the parabolic fits give also estimates of 
$\delta t_k$, which deviate to some extent from the direct estimates. 
For instance, for our typical sample of Fig.~\ref{fig_lnw} we find by 
direct estimate ($t$ in steps of 1~ns) 
\begin{equation} \label{1dmax} 
  \delta t_1 = 5~{\rm ns~~~and~~~}
  \delta t_2 = 6~{\rm ns}\,,
\end{equation}
while the parabolic fits give
\begin{eqnarray}  \label{fdmax} 
  \delta t_1 = (2.3\pm  9.9)\,{\rm ns}\,,~~~
  \delta t_2 = (3.7\pm 10.6)\,{\rm ns}\,,          
\end{eqnarray}
where, as in \cite{CNGS}, the error bars are standard deviations 
obtained from the parabolic fits under the assumption that the
center of the likelihood function is described by a Gaussian 
distribution. That the $\delta t_k$ values in (\ref{1dmax}) and 
(\ref{fdmax}) are both positive is an accident. E.g., for the sample 
of Fig.~\ref{fig_7lnw} one finds by direct estimate $\delta t_2=-6\,
$ns and $\delta t_2=(-4.7\pm 11.4)\, $ns from the parabolic fit.

The analysis of our 10$\,$000 samples allows to improve on the 
questionable Gaussian assumption. It turns out that the direct 
estimates of the $\delta t_k$ values from $\ln L^{\max}_k =\ln 
L_k(\delta t_k)$ are more robust than the estimates from parabolic 
fits, which can become unstable due to non-Gaussian behavior. 
One rare example is shown in Fig.~\ref{fig_161201lnw}. In that case a 
small $t$ shift, either to the left or to the right, does immediately 
shift one of the extrema $t_2^{\pi_1}$ or $t_2^{\pi_n}$ into regions 
of very small likelihood, so that a parabolic fit becomes impossible.

For a finite statistics the extraction of $\delta t_k$ from a given 
log-likelihood function $\ln L_k(\delta t)$ is a nonlinear procedure, 
so that we have to anticipate a bias \cite{Bbook}, which means
\begin{equation}
  \langle \delta t_k \rangle_n\ \ne\ 0\,.
\end{equation}
The exact maximum position, from which the deviation by a shift $t$ has 
to be calculated, is at $\langle \delta t_k\rangle_n$. This bias falls 
off with $1/n$, so that it tends to get swallowed by the $1/\sqrt{n}$ 
behavior of the statistical noise. For the direct $\delta t_k$ 
estimates we find in our models
\begin{eqnarray} \label{maxbias} 
  \langle \delta t_1 \rangle_{7612} &=& (-0.62\pm 0.10)\,{\rm ns}\,,
\\ \langle\delta t_2 \rangle_{7612} &=& (+0.22\pm 0.11)\,{\rm ns}\,,
\end{eqnarray}
which is in each case smaller than our discretization. 
The standard deviations were
\begin{equation} \label{sdv} 
  \triangle \delta t_1 = 9.8\,{\rm ns~~~and}~~~
  \triangle \delta t_2 = 10.2\,{\rm ns}\,.
\end{equation}
Combining both models gives 
\begin{equation} \label{ebar} 
  \triangle \delta t\ =\ 7.1\,{\rm ns}\,.
\end{equation}
To obtain estimates of these numbers from parabolic fits one has 
to eliminate 19 samples for which the fits are erratic. Afterwards,
averaging the standard deviations of the fits gives
\begin{eqnarray} 
  \triangle\delta t_1 &=&\ (9.7\pm 0.9)\,{\rm ns}\,\\ 
  \triangle\delta t_2 &=& (10.1\pm 1.1)\,{\rm ns}\,,
\end{eqnarray}
where the error of the error is with respect to estimates on a
single sample. Hence, for both models combined
\begin{equation} 
  \triangle \delta t\ =\ (7.0\pm 0.7)\,{\rm ns}\,.
\end{equation}
The $\triangle\delta t=7.5\,$ns value from the Gaussian fit to our 
typical sample is well consistent. Bias estimates stay much smaller 
than the standard deviation, but  differ from (\ref{maxbias}) as the 
fitting itself imposes a new non-linearity.  We abstain from giving 
these numbers, because they reflect also details of the fitting 
procedure like our definition of the central region.

\begin{figure}[tb] \begin{center} 
\epsfig{figure=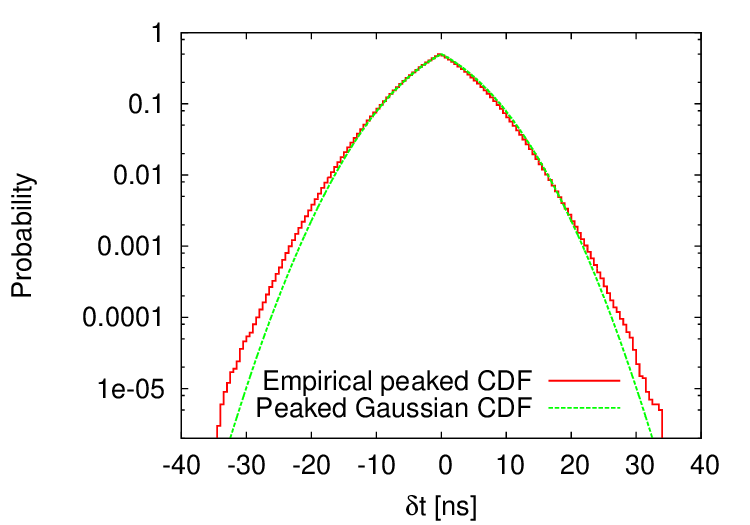,width=\columnwidth} 
\caption{Empirical peaked CDF of $\delta t$ estimates from 
$10^6$ samples (red online) in comparison with the peaked Gaussian 
CDF of standard deviation 7.1$\,$ns (green online). \label{fig_epCDF}} 
\end{center} \end{figure} 

As the distributions of the $\delta t_k$ are non-Gaussian, one 
may wonder whether Gaussian confidence limits provide a correct 
interpretation of the error bar (\ref{ebar}). Indeed, the empirical 
peaked CDF \cite{Bbook} found for the $\delta t$ estimates via the 
direct method is broader than the Gaussian peaked CDF with standard 
deviation 7.1$\,$ns. See Fig.~\ref{fig_epCDF}, which relies for the 
empirical peaked CDF on $10^6$ random times samples. The largest 
fluctuation in the negative direction is at $-35.5\,$ns which has 
thus a probability of approximately $10^{-6}$. This is still safely 
away from $-57.8\,$ns.

\begin{figure}[tb] \begin{center} 
\epsfig{figure=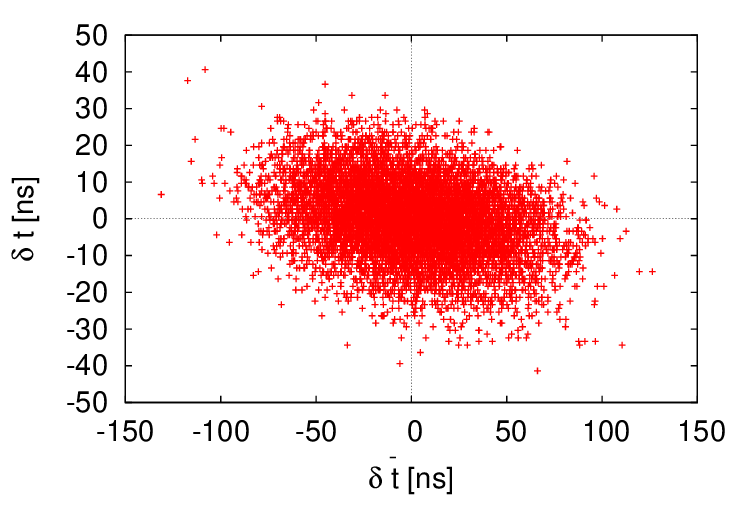,width=\columnwidth} 
\caption{Scatter plot of $(\delta\overline{t}^i_1,\delta t^i_1-{\rm 
bias})$, $i=1,\dots,n$, $N=10\,000$ (model~1). \label{fig_scatter}} 
\end{center} \end{figure} 

Finally in this section, we address correlations between the statistical
fluctuations of $\delta\overline{t}_k$ and those of $\delta t_k$. 
Using $i=1,\dots, N=10\,000$ samples, for sample $i$ we denote 
$\delta\overline{t}_k$ by $\delta\overline{t}^i_k$ and $\delta t_k$ 
by $\delta t^i_k$. They have expectation values 
\begin{equation}
  \langle\delta\overline{t}_k^i\rangle_{7612} = 
  \langle\delta t^i_k-{\rm bias}\rangle_{7612} = 0\,.
\end{equation}
The bias correction is so small that it can as well be ignored. 
Defining the correlation coefficients by
\begin{equation}
  C(\delta\overline{t}_k,\delta t_k)\ =\ \frac{ 
  \sum_{i=1}^n \delta\overline{t}^i_k\,(\delta t^i_k-{\rm bias}) }{
  \sum_{i=1}^n|\delta\overline{t}^i_k\,(\delta t^i_k-{\rm bias})|}
\end{equation}
their calculation from our $10\,000$ samples gives
\begin{eqnarray}
   \langle C(\delta\overline{t}_1,\delta t_1)\rangle_{7612} &=& -0.47\,,
\\ \langle C(\delta\overline{t}_2,\delta t_2)\rangle_{7612} &=& -0.43\,.
\end{eqnarray}
This, at the first glance, non-intuitive anti-correlation is for 
model~1 displayed in Fig.~\ref{fig_scatter}. It may be explained as
follows. When $\delta\overline{t}$ fluctuates to negative values, there 
is an overpopulation of small $t$ values in the sample. Therefore a 
shift in the same direction has a higher probability to produce 
large negative contributions to the log-likelihood function (\ref{lnLk}) 
than a shift in the opposite direction.

\section{Summary and Conclusions} \label{sec_sum}

We have investigated the problem of identifying a shift in a broad,
non-Gaussian distribution of data. For two model probability densities 
(PD) shown in Fig.~\ref{fig_2ppd} methods are developed and illustrated
to allow in samples of 7$\,$612 MC generated departure times to 
identify a time shift with a precision of about $\pm 7.1\,$ns (even
smaller than the error in our Eq.(\ref{deviation}) from 
Ref.~\cite{CNGS}). The noise of statistical fluctuations 
is with $\pm 23.8\,$ns much larger.

When the boundaries of the distribution are sufficiently sharp, uniform 
probabilities from binomials of the CDF provide excellent indicators.
Subsequently it has been demonstrated that the maximum likelihood method 
gives a quantitative estimate of the shift $\delta t$ and its statistical 
error, while bootstrap simulations allow one to avoid Gaussian assumptions.
Obviously, the analysis of this paper can be re-done for other desired
sample sizes.
\medskip 

\noindent {\bf Acknowledgments:}
I would like to thank Peter Hoeflich for useful discussions and for 
extracting Fig.~\ref{fig_2ppd} from Fig.~11 of Ref.~\cite{CNGS}. 
This work was in part supported by the DOE grant DE-FG02-13ER41942. 



\begin{thebibliography}{99}
\bibitem{MC} The sample used is just the first one encountered in
our simulations.
\end{thebibliography}
\end{document}